\documentstyle[12pt,psfig]{article}
\catcode`\@=11
\def\marginnote#1{}
\def\ifmath#1{\relax\ifmmode #1\else $#1$\fi}

\def\bold#1{\setbox0=\hbox{$#1$}%
     \kern-.025em\copy0\kern-\wd0
     \kern.05em\copy0\kern-\wd0
     \kern-.025em\raise.0433em\box0 }

\def\GENITEM#1;#2{\par\vskip6pt \hangafter=0 \hangindent=#1
   \Textindent{$ #2$ }\ignorespaces}

\newcount\hour
\newcount\minute
\newtoks\amorpm
\hour=\time\divide\hour by60
\minute=\time{\multiply\hour by60 \global\advance\minute by-
\hour}
\edef\standardtime{{\ifnum\hour<12 \global\amorpm={am}%
    \else\global\amorpm={pm}\advance\hour by-12 \fi
    \ifnum\hour=0 \hour=12 \fi
    \number\hour:\ifnum\minute<100\fi\number\minute\the\amorpm}}
\edef\militarytime{\number\hour:\ifnum\minute<100\fi\number\minute}
\def\draftlabel#1{{\@bsphack\if@filesw {\let\thepage\relax
  \xdef\@gtempa{\write\@auxout{\string
    \newlabel{#1}{{\@currentlabel}{\thepage}}}}}\@gtempa
    \if@nobreak \ifvmode\nobreak\fi\fi\fi\@esphack}
     \gdef\@eqnlabel{#1}}
\def\@eqnlabel{}
\def\@vacuum{}
\def\draftmarginnote#1{\marginpar{\raggedright\scriptsize\tt#1}}
\def\draft{\oddsidemargin -.5truein
        \def\@oddfoot{\sl preliminary draft \hfil
        \rm\thepage\hfil\sl\today\quad\militarytime}
        \let\@evenfoot\@oddfoot \overfullrule 3pt
        \let\label=\draftlabel
        \let\marginnote=\draftmarginnote

\def\@eqnnum{(\theequation)\rlap{\kern\marginparsep\tt\@eqnlabel}%
\global\let\@eqnlabel\@vacuum}  }
\def\preprint{\twocolumn\sloppy\flushbottom\parindent 1em
        \leftmargini 2em\leftmarginv .5em\leftmarginvi .5em
        \oddsidemargin -.5in    \evensidemargin -.5in
        \columnsep 15mm \footheight 0pt
        \textwidth 250mmin      \topmargin  -.4in
        \headheight 12pt \topskip .4in
        \textheight 175mm
        \footskip 0pt

\def\@oddhead{\thepage\hfil\addtocounter{page}{1}\thepage}
        \let\@evenhead\@oddhead \def\@oddfoot{} \def\@evenfoot{}
}
\def\titlepage{\@restonecolfalse\if@twocolumn\@restonecoltrue\onecolumn
     \else \newpage \fi \thispagestyle{empty}\c@page\z@
        \def\thefootnote{\fnsymbol{footnote}} }
\def\endtitlepage{\if@restonecol\twocolumn \else  \fi
        \def\thefootnote{\arabic{footnote}}
        \setcounter{footnote}{0}}  %\c@footnote\z@ }
\catcode`@=12
\relax
\def\be{\begin{equation}}
\def\ee{\end{equation}}
\def\bea{\begin{eqnarray}}
\def\eea{\end{eqnarray}}
\def\simlt{\stackrel{<}{{}_\sim}}

\def\NPB#1#2#3{{\it Nucl.~Phys.} {\bf{B#1}} (19#2) #3}
\def\PLB#1#2#3{{\it Phys.~Lett.} {\bf{B#1}} (19#2) #3}
\def\PRD#1#2#3{{\it Phys.~Rev.} {\bf{D#1}} (19#2) #3}
\def\PRL#1#2#3{{\it Phys.~Rev.~Lett.} {\bf{#1}} (19#2) #3}

\def\MPLA#1#2#3{{\it Mod.~Phys.~Lett.} {\bf#1} (19#2) #3}
\def\PR#1#2#3{{\it Phys.~Rep.} {\bf#1} (19#2) #3}

\def\HPA#1#2#3{{\it Helv.~Phys.~Acta} {\bf#1} (19#2) #3}
\def\JETPL#1#2#3{{\it JETP~Lett.} {\bf#1} (19#2) #3}

\def\bigint{{\displaystyle\int}}

\def\mst1{m_{\;\widetilde{t}_{1}}}

\def\mst2{m_{\;\widetilde{t}_{2}}}
\def\mst12{m_{\;\widetilde{t}_{1,2}}}

\def\msb1{m_{\;\widetilde{b}_{1}}}
\def\msb2{m_{\;\widetilde{b}_{2}}}
\def\msb12{m_{\;\widetilde{b}_{1,2}}}

\def\mtilde2{\widetilde{m}^{2}}

\relax
\begin{document}
\topmargin-2.5cm
%\draft
%\hoffset = .65 in

%\preprint
%
\begin{titlepage}
\begin{flushright}
IEM-FT-146/96 \\
hep--ph/9612212 \\
\end{flushright}
\vskip 0.3in
\begin{center}{\Large\bf SQUARKS AND SPHALERONS
\footnote{Work supported in part by the European Union
(contract CHRX/CT92-0004) and CICYT of Spain
(contract AEN95-0195).} }
\vskip .5in
{\bf J.M. Moreno, D.H. Oaknin} \\ 
\vspace{.2cm}
Instituto de Estructura de la Materia, CSIC, Serrano
123, 28006-Madrid, Spain

\vspace{.5cm}
and\\

\vspace{0.5cm}
 {\bf M. Quir\'os~\footnote{On leave from Instituto de Estructura 
de la Materia,
CSIC, Serrano 123, 28006-Madrid, Spain.}} \\
\vspace{.2cm} 
Theory Division, CERN, CH-1211 Geneva 23, Switzerland
\vskip.35in
\end{center}
\vskip2.3cm
\begin{center}
{\bf Abstract}
\end{center}

\begin{quote}
Electric charge and color breaking minima along third generation
squark directions do appear in the Minimal Supersymmetric 
Standard Model for particular regions of the corresponding 
supersymmetric parameters $A_t$, $\mu$, $\tan\beta$, 
$m_Q^2$ and $m_U^2$. We have studied possible instabilities of electroweak 
sphalerons along non-trivial squark configurations. We have found that
instabilities along the latter lie in the region of parameter
space where the electroweak minimum is not the global one.
Thus charge and color conservation imply that the standard electroweak
sphaleron is not destabilized along squark directions.

\end{quote}

\vskip3.cm

\begin{flushleft}
IEM-FT-146/96\\
November 1996 \\
\end{flushleft}
\end{titlepage}
\setcounter{footnote}{0}
\setcounter{page}{0}
\newpage
%
% BODY
\noindent 
{\bf 1.}
The existence of sphalerons (static and unstable solutions to
classical field equations) in the SU(2) gauge theory with Higgs 
fields in the fundamental representation has been 
known since long ago~\cite{sphalerons}, triggering the hope of generating
the baryon asymmetry of the Universe~\cite{baryogenesis} at the electroweak
phase transition~\cite{reviews}.
The value of the sphaleron energy, an essential ingredient
for baryogenesis mechanisms~\cite{Kuzmin1,Shapo1,McLerran1,Bochkarev}, 
has been the object of detailed
numerical calculations in the Standard Model~\cite{Manton,AKY,Yaffe}, 
and translates
into an upper bound on the Higgs boson mass well below the experimental
lower limit. This bound comes from the weakness of the phase
transition in the Standard Model.

The minimal supersymmetric extension of the Standard Model 
(MSSM)~\cite{susy} is
a very appealing candidate, both from the theoretical~\footnote{Yielding
a 'technical' solution to the hierarchy problem.} 
and the experimental~\footnote{LEP precision measurements leading to
gauge coupling unification.} side, to describe physics at the 
electroweak scale. The strength of the phase transition in the MSSM has been
extensively studied in the literature~\cite{earlyMSSM,mariano1,
mariano2}. Recently it has been
proven that the phase transition can be considerably strengthened in the
MSSM in the presence of light supersymmetric partners of the right-handed
top quark (stops) and small values of $\tan\beta$~\cite{CQW,JR,
Delepine}, opening thus the
window for baryogenesis at the electroweak phase transition. These
results have been recently confirmed by an explicit calculation of the
sphaleron solution and the sphaleron energy in the MSSM, at zero and
finite temperature, including all relevant one-loop radiative 
corrections~\cite{MOQ}.

The sphaleron solution presented in Ref.~\cite{MOQ} is the natural 
generalization of the Standard Model sphaleron solution and 
involves the two Higgs doublets and the SU(2) gauge fields. 
However, the main feature of supersymmetric models is the existence
of a plethora of new scalar fields, the sfermions, which might
affect the sphaleron solution if non-trivial configurations turn out
to be energetically favoured. In particular, for the region of 
supersymmetric parameters where the phase transition is strengthened,
there are electric charge and color breaking (CCB)~\cite{CCB}
local minima along the left-handed and/or right-handed stop
directions, which might destabilize the standard sphaleron in the MSSM
and then endanger previous results based on standard sphalerons and the
analysis of the phase transition. 

In this letter will study the effect of possible non-trivial
sfermion configurations on the sphaleron solution and its energy.
We find that as far as electric charge and color are not broken along
squark directions, i.e. as far as the standard electroweak minimum along
the neutral component of the Higgs fields remains as the global minimum,
the trivial (zero) configuration for sfermion fields is always 
energetically favoured. 

\vspace{1cm}
\noindent 
{\bf 2.} Neglecting sfermion mixing between different generations 
(as suggested
by constraints from flavour changing neutral current processes),
we can simplify the problem by considering just one generation
of squarks and sleptons.  On the other hand, as a first approximation,
we will fix to zero the slepton fields. 
The consistency of this fixing is protected by a global symmetry and will
be justified a posteriori by our results.
In fact we will work with third generation squarks, whose large Yukawa
couplings would allow tunneling~\cite{Claudson} 
from the symmetric vacuum at temperatures
above the electroweak phase transition temperature, or from the
standard non-symmetric electroweak vacuum, for temperatures below, to 
a possible lower CCB vacuum.
We will also work in the approximation
of taking $g_1=0$ so that the $U(1)_Y$ gauge field $B_{\mu}$ can be
consistently set to zero. 
This allows a spherically symmetric ansatz. 
So, we are looking for a sphaleron-like solution described by
\be
\label{viejos}
W_{\mu},
\; \; \; \; \;
H_1 = {\displaystyle  \left[
\begin{array}{c}
H_1^0 \\
H_1^-
\end{array}
\right]  },
\; \; \; \; \;
H_2  =   {\displaystyle \left[
\begin{array}{c}
H_2^+ \\
H_2^0
\end{array}
\right] }
\ee
and
\be
\label{nuevos}
\phantom{WW}
\;\;\;\;\;
Q_L = {\displaystyle  \left[
\begin{array}{c}
U_L \\
D_L
\end{array}
\right]  },
\; \; \; \; \;
U_R^c, \; \; \; \; \; D_R^c.
\ee

The lagrangian density for these fields is given by
\be
\label{lagrangian}
\begin{array}{rcl}
{\cal L} & = & -\frac{1}{4} W^a_{\mu\nu} W^{a\mu\nu}
+\left(D_{\mu}H_1\right)^{\dagger}\left(D^{\mu}H_1\right)
+\left(D_{\mu}H_2\right)^{\dagger}\left(D^{\mu}H_2\right)
+\left(D_{\mu}Q_L\right)^{\dagger}\left(D^{\mu}Q_L\right) \\ 
& & \\ 
& &
+\left(\partial_{\mu}U^c_R\right)^{\dagger}\left(\partial^{\mu}U^c_R\right)
+\left(\partial_{\mu}D^c_R\right)^{\dagger}\left(\partial^{\mu}D^c_R\right)
-V_{\rm eff}(H_1,H_2,Q_L,U^c_R,D^c_R)
\end{array}
\ee
where $W^{a}_{\mu\nu}$ stands for the $SU(2)$ field strength and 
$V_{\rm eff}$ is the effective potential.
We can expand the effective potential (at zero temperature) as
\be
\label{descomp}
V_{\rm eff}=V_0(H_1,H_2,Q_L,U^c_R,D^c_R)+V_1(H_1,H_2,Q_L,U^c_R,D^c_R)+\cdots
\ee
where $V_0$ is the tree-level potential and $V_1$ contains
the one-loop radiative corrections. The tree-level potential 
can be decomposed as:
\be
V_0 = V^D_{\rm SU(2)} + V^D_{\rm SU(3)} + V_{\em F} + V_{\rm soft}
\ee
with
\be
\begin{array}{rcl}
V^D_{\rm SU(2)} & = & \displaystyle \frac{g_2^2}{8}  
\left[ 
 (H_1^\dagger H_1)^2 + (H_2^\dagger H_2)^2 + (Q_L^\dagger Q_L)^2 \right. \\

&&  \\  & - &
2 \left( (H_1^\dagger H_1)(H_2^\dagger H_2) +
         (H_1^\dagger H_1)(Q_L^\dagger Q_L) +
         (H_2^\dagger H_2)(Q_L^\dagger Q_L) \right)  \\
&&  \\  & + & \left. 4 \left( 
|H_1^\dagger H_2|^2 + |H_1^\dagger Q_L|^2 +  |H_2^\dagger Q_L|^2  \right) 
\right]
\end{array}
\ee
and
\be
\begin{array}{rcl}
V^D_{\rm SU(3)} & = &\displaystyle \frac{g_3^2}{2}  
\left[   \displaystyle 
  \frac{1}{3}(Q_L^\dagger Q_L)^2 + \frac{1}{3}(U_R^{c \dagger} U^c_R)^2
+ \frac{1}{3}(D_R^{c \dagger} D^c_R)^2 \right. \\
&& \\
& - & (Q_L U^c_R) ^\dagger (Q_L U^c_R) + \displaystyle
    \frac{1}{3} (Q_L^\dagger Q_L)(U_R^{c \dagger} U^c_R) \\
&& \\
& - & (Q_L D^c_R) ^\dagger (Q_L D^c_R) + \displaystyle
\frac{1}{3} (Q_L^\dagger Q_L)(D_R^{c \dagger} D^c_R) \\
&& \\
& + & \left. (U^c_R D^c_R) ^\dagger (U^c_R D^c_R) - \displaystyle
\frac{1}{3} (U_R^{c \dagger} U^c_R)(D_R^{c \dagger} D^c_R)
\right]
\end{array}
\ee
the  SU(2) and SU(3) D-terms.
$V_{\em F}$ is determined from the superpotential
\be
W= h_t Q_L\cdot H_2 U_R^c+h_b Q_L\cdot H_1 D_R^c+\mu H_1\cdot H_2 ,
\ee
and reads:
\be
\begin{array}{rcl}
V_{\em F} & = &  
|h_b Q_L D^c_R - \mu H_2|^2  \, + \,   |h_t Q_L U^c_R + \mu H_1|^2 \\
& & \\  
&  +  &  |h_t H_2 U^c_R + h_b H_1 D^c_R|^2 
      \, + \,
     h_t^2 |Q_L\cdot H_2|^2  \, + \, h_b^2 |Q_L\cdot H_1|^2
\end{array}
\ee
The soft-breaking terms are:
\be
\label{soft}
\begin{array}{rl}
V_{\em soft} \; = & m_1^2\; H_1^{\dagger}H_1+
m_2^2\; H_2^{\dagger}H_2 + m_3^2\;(H_1\cdot H_2+h.c.)  \\  
&  \\
+ & m_Q^2 \; Q^{\dagger}_L Q_L + m_U^2 \; U^{c \dagger}_R U_R^c 
+   m_D^2 \; D^{c \dagger}_R D_R^c 
\\ & \\
+ & ( h_t A_t  \; Q_L \cdot H_2 U^c_R + 
      h_b A_b  \; Q_L \cdot H_1 D^c_R + h.c.)
\end{array}
\ee
where $m_1^2$, $m_2^2$ and $m_3^2$, can be traded in favour of the
supersymmetric parameters, $\tan\beta$ and $m_A$, using the minimization
conditions of the radiatively corrected effective potential.
Finally, for the one-loop effective potential one should use its expression    
in the $\overline{\rm DR}$~\cite{DR} renormalization scheme:
\be
\label{1loop}
V_1=\frac{1}{64\pi^2}{\rm Str}\left[M^4\left(\log\frac{M^2}{Q^2}
-\frac{3}{2}\right)\right]
\ee
where $Q$ is the renormalization scale. We have taken the approximation
 where only Higgs background fields are considered in
the one-loop correction, $V_1=V_1(H_1,H_2)$. 
This correction can be absorbed in a redefinition
of the tree-level parameters in the Higgs sector~\cite{CEQW} 
and hence provides large corrections to Higgs boson masses. On the other hand
the tree-level potential along the squark directions is dominated by the
strong $g_3$ and top-quark $h_t$ Yukawa couplings. Since the minima of the
potential will always be located at field values of the order of the
weak scale, we do not expect large radiative corrections for them 
and we can safely put to zero the background fields $Q_L$, $U_R^c$ 
and $D_R^c$ in (\ref{1loop}).

Choosing the temporal, radial gauge
we can write the static, spherically symmetric ansatz
\bea
\label{background}
W^a_j(\vec{x})&=&\frac{2 [1-f(r)]}{g_2\; r^2}\epsilon_{ajk}x_k \nonumber \\
H_1(\vec{x})&=&\widetilde{h}_1(r)\
\left[
\begin{array}{c}
1 \\
0
\end{array}
\right] \nonumber \\
H_2(\vec{x})&=&\widetilde{h}_2(r)\ 
\left[
\begin{array}{c}
0 \\
1
\end{array}
\right]   \\
Q_L(\vec{x})&=&\widetilde{q}(r)\ 
\left[
\begin{array}{c}
1 \\
0
\end{array}
\right] \nonumber \\
U^c_R(\vec{x})& =& \widetilde{u}(r) \nonumber \\
\nonumber
\eea
for the sphaleron~\footnote{Notice that we have fixed the right-handed squark
fields to the trivial configuration $D_L=D_R^c\equiv 0$. This result
is consistent with the smallness of the Yukawa coupling $h_b$, that
we will neglect from here on, and will be justified a
posteriori by the numerical results of this work.}, where 
$r^2=x^2+y^2+z^2$ and all functions are real. Notice that we have not
included gluons, while $Q_L$ and $U_R^c$ are aligned on
a unique SU(3) direction, which provides a self consistent ansatz.

Replacing the ansatz (\ref{background}) in (\ref{descomp}) we
obtain in a straightforward fashion the scalar potential as a function
of the background fields $\widetilde{h}_1$, $\widetilde{h}_2$, 
$\widetilde{q}$ and $\widetilde{u}$:
\be
\label{potCCB}
V_{\rm eff}=V_{\rm eff}(\widetilde{h}_1,\widetilde{h}_2,\widetilde{q},
\widetilde{u}). 
\ee
We have explored the region
where the supersymmetric parameters $$(m_Q,m_U,m_A,\tan\beta,A_t,\mu)$$
can take all values allowed by experimental and theoretical constraints.
In particular large values of $m_Q$ are preferred by LEP precision
measurements, while large values of $m_A$, $m_A\sim m_Q$, small
values of $\tan\beta$, $\tan\beta\simlt 3$, and moderately negative values
of $m_U^2$, $m_U^2>-m_t^2$, are favoured by the strength of the electroweak
phase transition, and will be focused on in the following discussion. 
This region has been recently proven to appear naturally from the
usual radiative scenario of electroweak symmetry breaking~\cite{Strumia}.
On the other hand it is clear that the presence of CCB minima 
and hence of possible instabilities of the standard sphaleron 
along the squark configurations will be highly favoured by $m_U^2<0$ values.

The analysis of the potential (\ref{potCCB}) reveals the 
presence of the standard electroweak minimum, at $\widetilde{q}=\widetilde{u}=0$, $\widetilde{h}_1=v\cos\beta$,
$\widetilde{h}_2=v\sin\beta$, where $v=174.1$ GeV is the standard Higgs vacuum 
expectation value (VEV), with a depth
\be
\label{profundidad}
V_{\rm ew}=-\frac{1}{4}m_H^2 v^2
\ee
where $m_H$ is the lightest Higgs boson mass in the limit $m_A\rightarrow
\infty$~\cite{MOQ}. For the case we are considering of large values of $m_A$,
$m_H$ is just the lightest Higgs boson mass.

For values of $\widetilde{A}_t\equiv A_t+\mu/\tan\beta<
\widetilde{A}_{t\ {\rm max}}$, with
\be
\label{amax}
\frac{\widetilde{A}_{t\ {\rm max}}^2}{m_Q^2}=1-\frac{g_3 m_H}
{\sqrt{3}h_t\sin\beta m_H}
\ee
there is a minimum at $\widetilde{h}_1=\widetilde{h}_2=\widetilde{q}=0$, 
${\displaystyle \widetilde{u}^2=\frac{3\widetilde{m}_U^2}{g_3^2} }$, 
where $\widetilde{m}_U^2\equiv -m_U^2$. The condition that
this CCB minimum not be lower than the standard minimum imposes on
$\widetilde{m}_U$ the upper bound $\widetilde{m}_U<
\widetilde{m}_U^{\rm crit}$ given by~\cite{CQW}
\be
\label{mucrit}
\widetilde{m}_U^{\rm crit}=\left(\frac{m_H^2 v^2 g_3^2}{6}\right)^{1/4}
\ee
We have plotted in Fig.~1 (long-dashed line) $\widetilde{m}_U^{\rm crit}$ 
given by Eq.~(\ref{mucrit}) as a function
of $\widetilde{A}_t$ for $m_Q=m_A= 500$ GeV and $\tan\beta=3$. We can
see the curve ends at a value of $\widetilde{A}_t$ given by 
Eq.~(\ref{amax}).

For $\widetilde{A}_t>\widetilde{A}_{t\ {\rm max}}$  
there is a CCB minimum with all fields
$\widetilde{h}_1$, $\widetilde{h}_2$, $\widetilde{q}$ and $\widetilde{u}$ 
acquiring VEVs. The location of the minimum
can only be computed numerically and $\widetilde{m}_U^{\rm crit}$, at which
the CCB minimum becomes degenerate with the standard electroweak minimum is
shown in Fig.~1 (short-dashed line) for the same values of the supersymmetric
parameters.

In short, in the region below (above) the dashed line the electroweak minimum
(the CCB minimum) is the global one, and only below the dashed line is
electric charge and color guaranteed to be unbroken. For illustration we 
have depicted also in Fig.~1 the (thin solid) line where the right-handed
squarks are massless. Then only below the thin solid line there exists
a (local or global) standard electroweak minimum.

\vspace{1cm}
\noindent
{\bf 3.} The last issue we want to address is the possible (in)stability
of the standard sphalerons in the MSSM~\cite{MOQ}. It is again clear that
the region $m_U^2<0$, where the right-handed squark direction is 
unstable at the origin,
favours the presence of such instabilities.
One should now substitute the ansatz~(\ref{background}) into the 
Euler-Lagrange equations, 
\be
\begin{array}{rcl}
( D_\nu W^{\mu\nu})^a &=& 
 i {\displaystyle\frac{g_2}{2}\sum_{Y=H_{1,2},Q_L} \left[
 Y^\dagger \sigma^a (D^\mu Y) - (D^\mu Y)^\dagger \sigma^a Y\right]  }\\
&& \\
{\displaystyle D_\mu D^\mu X}&=&-{\displaystyle  
\frac{\partial V}{\partial X^\dagger} }
\end{array}
\label{ecuaciones}
\ee
where $X = H_{1,2},Q_L,U^c_R,D^c_R$, supplied by the appropriate
boundary conditions. In some cases we expect to end up with
a solution described by vanishing squark fields, 
whereas in other situations non-vanishing profiles 
can be preferred. Before embarking ourselves in the complete
solution of the equations of motion, we will study the
possible standard sphaleron instabilities along the squark field
configurations. Hence we will treat the standard ({\em squarkless})
sphaleron as a background and we will study the corresponding squark 
equations in this background. This approximation will be justified a 
posteriori, and will provide us with an intuitive picture of the 
parameter space region where squarks are involved in the sphaleron 
solution. 

The energy of the squark fields in the
sphaleron background is given by:

\bea
E_{\widetilde q}& =&
4\pi\; \bigint dr\; r^2 
\left[ \phantom{\frac{1}{2}}\right.
      ( \partial_r \widetilde{q\;})^2 + \frac{2}{ r^2}(1-f)^2 
\widetilde{q\;}^2 
+ (\partial_r \widetilde{u\;})^2  \nonumber \\  
     & +& \frac{g_2^2}{8} \left[\widetilde{q\;}^4 +2 \widetilde{q\;}^2 
(\widetilde{h}_1^2 
- \widetilde{h}_2^2 )\right] 
      + \frac{g_3^2}{6}(\widetilde{q\;}^2-\widetilde{u\;}^2)^2   \\
& + &  
        h_t^2 \widetilde{q\;}^2 \widetilde{u\;}^2 + 2 \mu h_t 
\widetilde{q\;}\; \widetilde{u\;} \widetilde{h}_1  
      +  h_t^2 \widetilde{h}_2^2 \widetilde{u\;}^2 + h_t^2 
\widetilde{h}_2^2 \widetilde{q\;}^2 \nonumber\\ 
     & +&\left.  m_Q^2 \widetilde{q\;}^2 + m_U^2 \widetilde{u\;}^2 + 2 h_t A_t 
\widetilde{q\;} \widetilde{h}_2 \widetilde{u\;}  \phantom{\frac{1}{2}}
\right] \nonumber
\eea

Non-vanishing squark field configurations will be associated to negative
values of the previous functional. We can split it as follows:
\be
E_{\widetilde q} = \delta^2 E_{\widetilde q} + \delta^4 E_{\widetilde q}
\ee
where the first piece includes quadratic terms in the squark fields 
and the second one collects the remaining, quartic, ones. Notice that
$\delta^4 E_{\widetilde q} \geq 0 $ and then a decreasing of the energy
is possible only in the case where the quadratic terms are negative.
These are given by:
\be
\label{curvature}
\delta^2 E_{\widetilde q} = 
4\pi\; \bigint dr \left[
\begin{array}{cc}
r \widetilde{q}(r)& r \widetilde{u}(r)
\end{array}\right]
 {\cal M}^2 
\left[
\begin{array}{c}
r \widetilde{q}(r) \\
r \widetilde{u}(r)
\end{array}\right]
\ee
where
\bea 
\label{operador}
{\cal M}^2 = 
\left[  \begin{array}{cc}
{\displaystyle -\partial^2_r + \frac{2}{r^2}(1-f)^2  +
\frac{g_2^2}{4}( \widetilde{h}_1^2 - \widetilde{h}_2^2)  
+ h_t^2 \widetilde{h}_2^2 + m_Q^2 }  
&  h_t (\mu \widetilde{h}_1 + A_t  \widetilde{h}_2) \\
& \\
h_t (\mu \widetilde{h}_1 + A_t \widetilde{h}_2) &
-\partial^2_r + h_t^2 \widetilde{h}_2^2 + m_U^2  
\end{array} \right]
\eea 
To determine the instabilities of the standard sphaleron along
the squark direction one must diagonalize the quadratic operator in
(\ref{curvature}). Each eigenvector of ${\cal M}^2$ with a\
negative eigenvalue
\be
\label{negativo}
{\cal M}^2
\left[
\begin{array}{c}
r \widetilde{q}(r) \\
r \widetilde{u}(r)
\end{array}
\right]
=-\omega^2
\left[
\begin{array}{c}
r \widetilde{q}(r) \\
r \widetilde{u}(r)
\end{array}
\right]
\ee
generates an instability.

We have studied the spectrum of the operator (\ref{operador})
numerically, using a discretized representation of 
the quadratic energy functional. 
The MSSM sphaleron is characterized 
by just one scale $\sim M_W$, since it is mainly controlled
by the $W$ mass and the lightest Higgs boson mass, the latter never
being hierarchically larger than the former~\footnote{This situation is 
different in the Standard Model, where the quartic coupling of the
Higgs field is, in principle, arbitrary, so that it can be
arbitrarily heavier than the $W$ boson, in which case a second
scale, $M_H$, appears.}. 
When squarks are included, a second scale, $M$,
could emerge in our solution. It would be
roughly given by the inverse radius of the region where
squarks are non-vanishing. The value of this scale
depends on the spectrum of our MSSM model, but we
expect it to be in the range  $M_W\simlt M\simlt M_{\rm SUSY}$, where
$M_{\rm SUSY}$ is typically $\simlt 1 $ TeV. In order to
cover both scales, we will follow the approach of Ref.~\cite{Yaffe} and use 
a uniform discretization of the variable:
\begin{equation}
s = \ln \left[ \frac{1 + M r}{ 1 + M_W r} \right]
     /
    \ln (M/M_W)
\end{equation}
where $s$ goes from 0 to 1 when $r$ varies from  0 to
infinity. We have done our analysis using 200 points
and taking  $M$ values within the 
above mentioned range.  We have verified, by increasing the
number of points and comparing, that the lower eigenvalue 
is calculated with and error smaller than
0.001 in $M_W^2$ units.

We have plotted in Fig.~1, for the same values of the supersymmetric
parameters $(\tan\beta,m_A,m_Q$) as before the (thick solid)
line along which there
appears a negative eigenvalue along a non-trivial squark configuration.
So above (below) the thick solid line the standard sphaleron in the
MSSM is unstable (stable). We can see from Fig.~1 that the instability
appears only in the region where the CCB minimum is the global one and
electric charge and color are broken.
We have stopped the instability line when it intersects the thin solid
line, where the standard electroweak minimum disappears. We have not shown
the region with $m_U^2>0$, but it becomes clear from the behaviour of the
thick solid line in Fig.~1 that there is no instability region  for
$m_U^2>0$ and any value of $\widetilde{A}_t$.

We have also explored general values of the supersymmetric parameters in
the region $1\leq\tan\beta\simlt 20$ (where we can comfortably neglect
the bottom Yukawa coupling $h_b$ as compared to $h_t$), 
100 GeV $\simlt m_A\simlt m_Q$ and
200 GeV $\simlt m_Q\simlt 500$ GeV, and we have found a qualitative
behaviour similar to that in Fig.~1. Even in the region of 
large values of $\tan\beta$,
where $h_b$ can no longer be neglected, our results holds since the possible
instability along $D_L,D_R^c$ would be controlled by the entry
$$-\partial_r^2+h_b^2 \widetilde{h}_1^2+m_D^2$$ in the corresponding
quadratic operator, similar to that in (\ref{operador}), which in turns 
depends on the value of the bottom-quark mass. Therefore, the present
experimental bounds on supersymmetric masses impose $m_D^2>0$ which prevents,
using the results obtained in this paper, the existence of instabilities
along the squark fields $D_L,D_R^c$. Since the previous results 
obviously apply to slepton fields, our choice of the ansatz where slepton
and down-squark fields vanish remains fully justified.

\vspace{1cm}\noindent
{\bf 4.}
Using the previous results, which concern the MSSM standard 
electroweak sphaleron at
zero temperature, we can infer the stability of the standard sphaleron
against non-trivial squark configurations
at temperatures higher than, or equal to, the critical temperature
of the electroweak phase transition, $T_c \sim$ 100 GeV. The argument goes 
as follows: The main effect of finite temperature corrections in the
MSSM sphaleron energy and profiles can be encoded, to a reasonable
accuracy, in the (asymptotic) vacuum expectation values of the Higgs
fields, i.e. $v_1(T)$ and $v_2(T)$~\cite{MOQ}. 
On the other hand, the only modification
in the quadratic operator (\ref{operador}) from finite temperature 
corrections is to replace the squark masses $m_Q^2$, $m_U^2$ by effective
thermal masses, $m_Q^2(T)=m_Q^2+\Pi_L(T)$ and $m_U^2(T)=m_U^2+\Pi_R(T)$, 
where the self-energies $\Pi_L(T)$ and $\Pi_R(T)$ can be found in 
Ref.~\cite{mariano2}, and whose precise expressions are not important for
the sake of the present discussion. 
A necessary condition at $T\geq T_c$ is that
$m_Q^2(T)>0$, $m_U^2(T)>0$ since otherwise the phase transition would
proceed at some temperature $T_{\rm CCB}\geq T_c$ towards the 
finite temperature CCB
minimum through a strong first order phase transition~\cite{CQW}. Using the
results of the present work we can infer there would be 
no instability region for the range of temperatures $T\geq T_c$, since
negative squared masses have been found to be a necessary condition for the
existence of instabilities. In other words, the stability
of the standard electroweak minimum at the critical temperature 
implies stability of the standard electroweak sphaleron along squarks
configurations. Of course,
below $T_c$, $m_Q^2(T)$ and/or $m_U^2(T)$ can become negative and 
an instability region will start to grow up and evolve towards the zero
temperature result shown in Fig.~1.

\vspace{1cm}\noindent
{\bf 5.} 
In conclusion, we have studied possible instabilities of the standard 
electroweak sphaleron in the MSSM along the most dangerous third 
generation sfermion
configurations, with CCB minima, which might decrease the sphaleron
energy. At zero temperature, we have found that 
instabilities lie in the region of 
supersymmetric parameters where CCB minima are the global minima.
Thus imposing electric charge and color conservation is a sufficient
condition to ensure the stability of the standard sphaleron along 
third generation sfermion directions. 
Finally, at the electroweak phase transition
temperature, imposing the stability of the electroweak
minimum prevents the appearance of instabilities along sfermions
configurations.

\vspace{1cm}
{\large\bf Acknowledgements}

\vspace{0.5cm}
\noindent
We thank M. Shaposhnikov for useful comments.

%\newpage

%\newpage
%%%%%%%%%%%%%%%%%%%%%%%%figure%%%%%%%%%%%%%%%%%%%%%%%%
%\phantom{.}
\begin{figure}[b]
%\psdraft
\centerline{
\psfig{figure=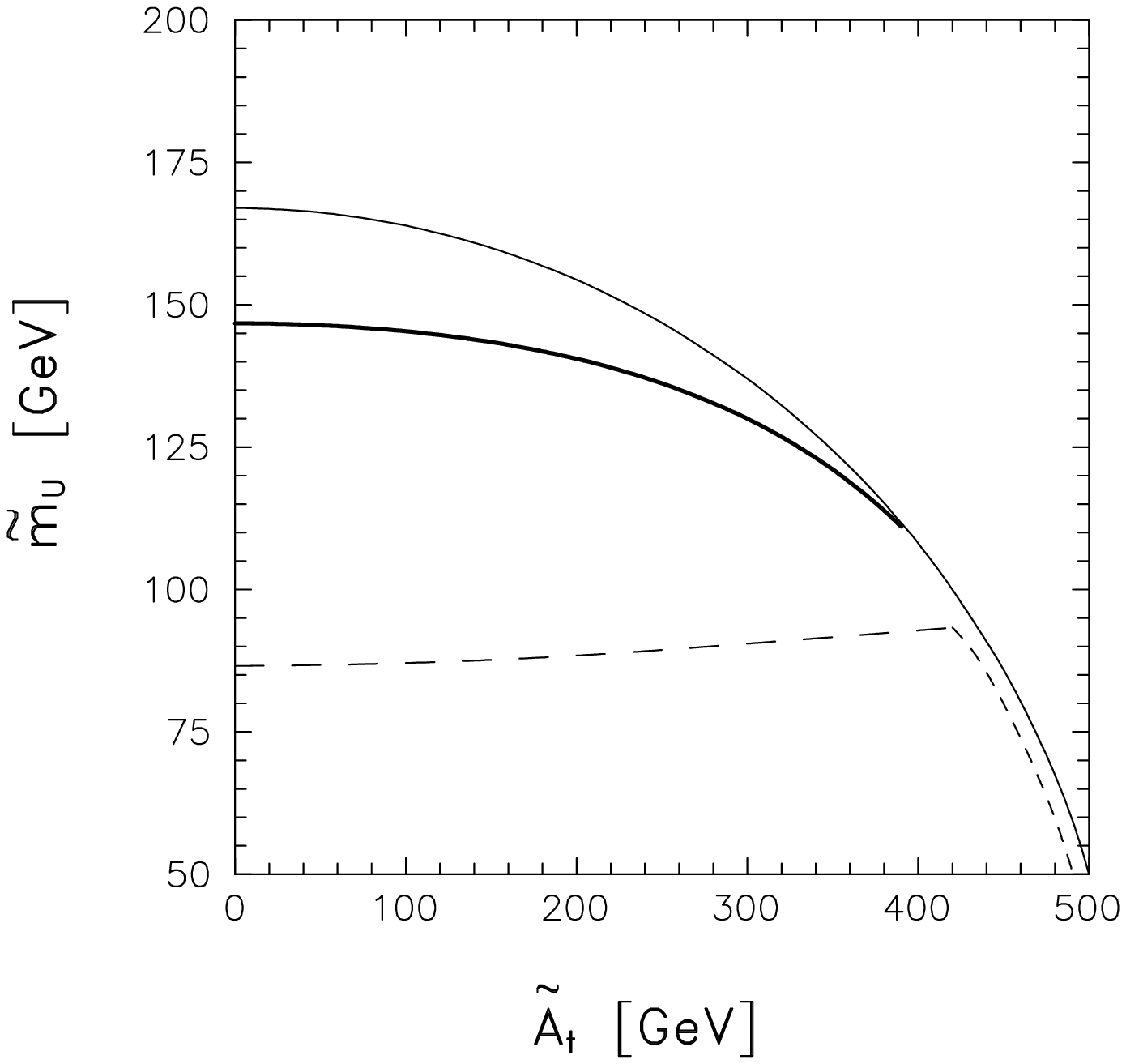,height=13cm,bbllx=4.5cm,bblly=1.cm,bburx=14.cm,bbury=14.cm}}
\caption{For $\tan\beta=3$ and $m_Q=m_A=500$ GeV: a) Above the thin
solid line the electroweak minimum disappears; b) Above (below) the
dashed lines electric charge and color are  broken (conserved); c) Above
(below) the thick solid line the standard sphaleron is unstable (stable)
along the squark directions.}
\label{f1}
\end{figure}
%%%%%%%%%%%
\end{document}